\documentclass[preprint,12pt]{elsarticle}

\usepackage{amssymb}
\usepackage{color}
\usepackage{amsbsy}
\usepackage{amsmath}

\usepackage[italian, english]{babel}
\usepackage{amsmath}
\usepackage{amssymb}
\usepackage{amsfonts}
\usepackage{subfig}
\usepackage{latexsym}
\usepackage{color}
\usepackage{fleqn}
\usepackage{amscd}
\usepackage{verbatim}

\usepackage[T1]{fontenc}
\usepackage{amsthm}
\usepackage{mathrsfs}
\usepackage{amssymb}
\usepackage[a4paper,vmargin={33mm,33mm},hmargin={33mm,33mm}]{geometry}

\theoremstyle{plain}

\providecommand*{\de}{\ensuremath{\mathrm{d}}}

\providecommand*{\eu}{\ensuremath{\mathrm{e}}}

\providecommand*{\x}{\ensuremath{\boldsymbol{x}}}

\providecommand*{\uf}{\ensuremath{\boldsymbol{u}}}

\providecommand*{\deti}{\ensuremath{\partial_t}}

\providecommand*{\Id}{\ensuremath{\mathbb{I}}}

\numberwithin{equation}{section}
\numberwithin{theorem}{section}
\numberwithin{thm}{section}
\numberwithin{example}{section}
\numberwithin{prop}{section}
\numberwithin{oss}{section}

\journal{Comptes Rendus Physique}

\begin{document}

\begin{frontmatter}

%% Title, authors and addresses

%% use the tnoteref command within \title for footnotes;
%% use the tnotetext command for theassociated footnote;
%% use the fnref command within \author or \address for footnotes;
%% use the fntext command for theassociated footnote;
%% use the corref command within \author for corresponding author footnotes;
%% use the cortext command for theassociated footnote;
%% use the ead command for the email address,
%% and the form \ead[url] for the home page:
%% \title{Title\tnoteref{label1}}
%% \tnotetext[label1]{}
%% \author{Name\corref{cor1}\fnref{label2}}
%% \ead{email address}
%% \ead[url]{home page}
%% \fntext[label2]{}
%% \cortext[cor1]{}
%% \address{Address\fnref{label3}}
%% \fntext[label3]{}

\title{Motility-induced phase separation and coarsening in active matter}

%% use optional labels to link authors explicitly to addresses:
%% \author[label1,label2]{}
%% \address[label1]{}
%% \address[label2]{}

\author[label1]{Giuseppe Gonnella}
\ead{gonnella@ba.infn.it}

\author[label2]{Davide Marenduzzo}
\ead{dmarendu@ph.ed.ac.uk}

\author[label3]{Antonio Suma}
\ead{antonio.suma@gmail.com}

\author[label4]{Adriano Tiribocchi}
\ead{adriano.tiribocchi@pd.infn.it}

\address[label1]{Dipartimento di Fisica, Universit\`a di Bari {\rm and}  \\
INFN, Sezione di Bari, via Amendola 173, Bari, I-70126,
Italy}

\address[label2]{SUPA, School of Physics, University of Edinburgh - Edinburgh EH9 3JZ, UK}

\address[label3]{SISSA - Scuola Internazionale Superiore di Studi Avanzati,\\
Via Bonomea 265, 34136 Trieste 
Italy}

\address[label4]{Dipartimento di Fisica e Astronomia, Universit\`a di Padova   \\
via Marzolo 8, Bari, I-70126, Italy}

\begin{abstract}

Active systems, or active matter, are self-driven systems which live, or 
function, far from equilibrium -- a paradigmatic example which we
focus on here is provided by a suspension of self-motile particles. 
Active systems are far from equilibrium because their microscopic constituents 
constantly consume energy from the environment in order to do work, for 
instance to propel themselves.
The nonequilibrium nature of active matter leads to a variety of non-trivial
intriguing phenomena. An important one which has recently been the subject of
intense interest among biological and soft matter physicists is that of the 
so-called "motility-induced phase separation", whereby self-propelled 
particles accumulate into clusters in the absence of any explicit attractive
interactions between them. Here we review the physics of motility-induced
phase separation, and discuss this phenomenon within the framework of the 
classic physics of phase separation and coarsening. We also discuss theories for  bacterial
colonies   where coarsening may be arrested. Most of this work will focus on the case
of run-and-tumble and active Brownian particles in the absence of 
solvent-mediated hydrodynamic interactions -- we will briefly discuss 
at the end their role, which is not currently fully understood in this context.

\end{abstract}

\begin{keyword}
%% keywords here, in the form: 
active matter \sep phase separation 

%% PACS codes here, in the form: \PACS code \sep code

%% MSC codes here, in the form: \MSC code \sep code
%% or \MSC[2008] code \sep code (2000 is the default)

\end{keyword}

\end{frontmatter}

%% \linenumbers

\section{Introduction}
\label{sec:introduction}

Active systems are self-driven systems which live, or function, far from thermodynamic equilibrium. A paradigmatic example is provided by a suspension of interacting self-motile particles in equilibrium with a thermal bath at temperature $T$. This system, like active matter in general, is characterized by the continuous  conversion of  internal energy into work or movement~\cite{vicsek,ebeling}, which drives it far from equilibrium even in steady state~\cite{Ramaswamy10}.

Nature offers many examples of active matter, at very different scales: these range, for instance, from the cytoskeleton of eukaryotic cells, to whole bacterial colonies and  algae suspensions, and even to bird flocks and schools~\cite{Toner05,Fletcher09,Menon10,Ramaswamy10,Cates12,Vicsek12,Marchetti13,Marenduzzo14,Elgeti15}. 
Self-propelled units can also be artificially realized in the lab in many different ways, for example by anisotropic surface treatment of colloidal particles~\cite{Walther,Paxton04,Paxton06,Hong06,Howse07,Golestanian07,Palacci10}. 

The inherently nonequilibrium nature of active matter endows it with a number of non-trivial features that have no analogue in passive, equilibrium  materials, 
such as a suspension of Brownian colloidal particles. 
For example, bacterial swimmers may accumulate near walls due to activity alone~\cite{Elgeti09,Lauga13}, and their large scale diffusive motion may be rectified by using funnels or other asymmetric geometrical obstacles~\cite{Tailleur09,Chaikin07}. Within a bacterial suspension, it is also well known both that there are violations of the fluctuation-dissipation theorem~\cite{Chen07}, and that the diffusion of passive tracers is dramatically affected by the presence of active swimmers~\cite{wu,Palacci10,Leptos09,Kasyap14,Pushkin14,Morozov14,valeriani2011colloids}.
The existence of a parameter to be interpreted as an effective temperature~\cite{cugl:review} for active matter has been analysed 
in~\cite{cugl-mossa1,cugl-mossa2,cugl-mossa3,Palacci10,Suma14b}.
The interplay between activity and the rodlike shape of bacteria and most other active particles leads to further surprising phenomena, such as the large scale coherent turbulent-like motion observed in bacterial
 monolayers~\cite{Mendelson99,wu,Dombrowski04,Hernandez05,Riedel05,Sokolov07,Zhang09}, or the strikingly non-Newtonian rheology found in simulations of model active gels~~\cite{Liverpool06,Cates08,Giomi08,Foffano12,Fielding10}.

Our main focus in this review is another example of a phenomenon that is unique to active systems, the so-called ``motility-induced phase separation'', which has attracted a lot of attention recently~\cite{Tailleur08,Fily12,Fily14,Redner13,Stenhammar13,Suma13,Suma14,Levis-Berthier,Stenhammar14}. This term refers to the phase separation into a concentrated and a gas-like dilute phase observed in a system of self-propelled particles interacting solely via steric or excluded volume repulsion, 
when the overall density is large enough~\footnote{We will not consider in this review models, like the 
Vicsek model~\cite{vicsek}, 
 where self-propelled particles interact via explicit alignement interaction and phase transition is related 
to spontaneous symmetry breaking
  - see also  the end of 
Section 3.3.}.
The physics of motility-induced phase separation (henceforth MIPS) can be understood as follows. Imagine that due to a fluctuation the local density of self-propelled particle increases in some part of the system. The particles in that region will therefore slow down due to the enhanced crowding. Now, as we will understand more formally in Section 2, self-motile particles (such as bacteria) accumulate where they move more slowly, much as pedestrians in a busy street (and very much unlike colloidal particle undergoing Brownian motion).
This potentially triggers a positive feedback loop, whereby particles accumulate where they are slower, further slow down due to the crowding, then accumulate even more etc. The theory which we present in Section 2, and which is fully confirmed by Brownian dynamics simulations of self-propelled spherical particles (discussed in Section 3), predicts that these clusters should coarsen so as to lead to complete phase separation. Indeed, the theory suggests that this nonequilibrium phase separation should be very similar (in the same universality class) to the phase separation in a system of passive colloidal particles which are subject to a mutual attractive interaction -- according to this view the main role of activity is to provide an effective attraction between particles, in the absence of any thermodynamic attractive force. Indeed the macroscopic description of MIPS is very similar to a Cahn-Hilliard equation for a phase separating binary fluid, albeit with some subtle important differences.

Recently, experiments have also been reported where MIPS was observed in suspensions of self-motile particles only subject to steric interactions~\cite{Buttinoni13b}. Clustering of self-propelled synthetic particles was also found in Ref.~\cite{Palacci10a,Palacci13,Palacci14}, although in both those cases the coarsening appeared to arrest, for reasons which are still not fully understood. In particular, it is not clear to date to what extent the theory of MIPS (as discussed above and in Section 2) applies to the work of Ref.~\cite{Palacci10a,Palacci13,Palacci14}, or how to modify it in an experimentally relevant way so as to lead to microphase separation, or arrested coarsening. 

Another important aspect of MIPS worth discussing is that the theory normally considers spherical self-propelled particles, whereas in most cases active particles are in practice elongated (think for instance of bacteria or chemically powered nanorods). It is therefore of interest how MIPS appears in a suspensions of rodlike, or elongated, rather than spherical, microswimmers. While several simulations of rodlike active particle exists~\cite{Peruani12prl,Baskaran12,Wensink12}, it is only recently that the effect of particle shape on MIPS has been directly addressed -- for instance with simulations of dumbbells~\cite{Suma14}. We will review these works in Section 3. Section 3 also contains a brief description of coarse grained theories for self-propelled particles which align locally (for instance due to steric interactions if they are rodlike in shape); such theories are more complicated that the Cahn-Hilliard-like equation governing the evolution of active spherical particles because they need to follow the dynamics of the polarisation field as well.

Finally, we will consider other ways in which MIPS can be extended in Section 4. First, we will discuss an intriguing and generic way to arrest phase separation, by coupling the motility-induced feedback described above to a logistic law for the population of active particles, used as a simple metaphor for bacterial reproduction~\cite{Cates10}. The arrested patterns predicted by the resulting theory closely resemble those found experimentally in growing bacterial colonies within semi-solid media such as agarose gel, which are normally explained instead by bacterial chemotaxis~\cite{Murray}. Second, all theories discussed so far neglect solvent-mediated hydrodynamic interactions between active particles. However, in a nonequilibrium active suspension, hydrodynamic can potentially modify the steady state attained by the system, and not only its transient dynamics. Furthermore, unlike active particles in Brownian simulations, neutrally buoyant hydrodynamic swimmers in a solvent need to be force free, so that the minimal active contribution is a force dipole~\cite{Ramaswamy10}. In Section 4.2 we will briefly review some of the studies which have appeared in the literature for suspensions of spherical squirmers~\cite{Evans11,Alarcon13,Fielding14,Zottl14}, where the activity arises through a non-zero imposed slip velocity at their boundary. The consequence of solvent-mediated hydrodynamics for MIPS is currently unclear, and some conflicting conclusions have been reached in the literature, possibly reflecting the dependence on details of near field interactions. 

\section{The physics of motility: self-trapping of self-propelled particles}
\label{motility}

In this Section we provide a simple theoretical framework to understand the theory of MIPS, or self-trapping of self-propelled particles~\cite{Cates12,Tailleur08,Stenhammar13,Tailleur09}.
In particular, we will describe this phenomenon in terms of an instability appearing in the coarse-grained ``hydrodynamic'' equation manifested by the presence of a negative diffusion coefficient. 
To establish a hydrodynamic equation for the self-propelled swimmers, we use here a kinetic approach; similar calculations have been also presented in Ref.~\cite{Marenduzzo14} and are 
an adaptation of Refs.~\cite{Schnitzer93,Tailleur10}. 

\subsection{A coarse-grained theory of motility-induced phase separation}

Let us  consider  a population of  swimmers that runs for a time $\tau$ with velocity $v$ between two successive
 tumbles (sudden
changes of directions~\cite{Cates12}). 
Let $f (\x, \uf, t)$ be the distribution of active particles with
position $\x$, swimming in the direction $\uf$ at the time $t$.
The kinetic (Boltzmann-like) equation for the evolution of $f$ in $d$ 
dimensions can be written as
\begin{equation}
\label{eq:swimmerskinetic}
\deti f(\x,\uf,t) = -\nabla_{\x} \cdot \left[v\uf f(\x,\uf,t)\right] - \frac{ f(\x,\uf,t)}{\tau} + \frac{\int  f(\x,\uf,t) \de^{d-1}\uf}{\tau\underbrace{\int \de^{d-1}\uf}_{\textrm{solid angle }\Omega_d}} .
\end{equation}
The first term on the right hand side of the equation is a convective term, where $v\uf$ is the
velocity; note that in general we are interested in the case in which $v$ depends
on the spatial position, $\x$. The second term is a linear term accounting for the particle that
"tumble out" of $\uf$, with a rate $1/\tau$ ; the same rate appears in the third term, that accounts for
the particles "tumbling into" the direction $\uf$.
The integral of $f$ appearing in the third term is the zero-th order moment of $f$ with respect to
the orientation $\uf$, which is nothing but the density $\rho$ of swimmers in $\x$. Finally, $\Omega_d$ denotes the $d$-dimensional solid angle.

A way to analyze Eq.~(\ref{eq:swimmerskinetic}) is to consider the so-called hydrodynamic limit
%, also discussed above, in
in which suitable moments of the distribution functions, which are linked to macroscopic observables, 
change smoothly in space and in time. The idea is
 %then, in practice, 
to perform a gradient expansion (which is very similar to the so-called Chapman-Enskog expansion~\cite{chapman})
of Eq.~(\ref{eq:swimmerskinetic}), which assumes that successive derivatives (gradients)
are smaller and smaller on macroscopic (coarse grained scales). The distribution functions is then
written as
\begin{equation}
\label{eq:C_E_series}
f= f^{(0)}+ f^{(1)}+ f^{(2)}+...
\end{equation}
where $f^{(n)}$ is a contribution which is $\sim {\mathcal O}{\left(D^n\right)}$, where $D$ denotes derivatives over
space or time. We assume that in the hydrodynamic limit
 every successive derivative applied to $f$ makes it much smaller, that is:
\[
\mathcal{O}\left(D^n\right) \ll \mathcal{O}\left(D^{n-1}\right) \ll ...\ll \mathcal{O}\left(D^1\right) \ll \mathcal{O}\left(D^0\right) \equiv \Id,
\]
hence
\begin{equation}
f^{(n)}\ll f^{(n-1)} \ll .. \ll f^{(1)} \ll f^{(0)}. 
\end{equation}
Then, we can set up a hierarchy of equations at various orders in $D$,
by equating terms of order $D^n$ in the left and right hand side of Eq.~(\ref{eq:swimmerskinetic}),
as follows:
\begin{equation}
\label{eq:C_E_zeroth}
\mathcal{O}(D^0):\qquad f^{(0)}= \frac{\rho}{\Omega_d}
\end{equation}
\begin{equation}
\label{eq:C_E_first}
\mathcal{O}(D^1):\qquad \deti f^{(0)} + \nabla_{\x} \left(v\uf f^{(0)}\right)= -\frac{f^{(1)}}{\tau} 
\end{equation}
\begin{equation}
\label{eq:C_E_second}
\mathcal{O}(D^2):\qquad \deti f^{(1)} + \nabla_{\x} \left(v\uf f^{(1)}\right)= -\frac{f^{(2)}}{\tau} .
\end{equation}

From Eq.~(\ref{eq:C_E_zeroth}) we find that no term other than $f^{ (0)}$ contributes to the coarse-grained
hydrodynamic density $\rho$. Then,  by summing Eqs.~(\ref{eq:C_E_first}) and (\ref{eq:C_E_second}),
and taking the integral over the directions we get
\begin{equation}
\label{eq:C_E_firstandsecond_order_integrated}
\deti \rho + \int  \partial_\beta \left(vu_\beta f^{(0)}\right) \de^{d-1}\uf
 +  \int  \partial_\beta \left(vu_\beta f^{(1)}\right)\de^{d-1}\uf = 0 , 
\end{equation}
where the relations $\int f^{(1)}(\x,\uf,t)\de^{d-1}\uf = \int f^{(2)}(\x,\uf,t)\de^{d-1}\uf = 0 $ have been used
(Greek indices denote Cartesian components, and the usual convention
of summation over repeated indices is implied).
Then we can use   the expressions  of $f^{(0)}$ from Eq.~(\ref{eq:C_E_zeroth}) and 
$f^{(1)}$ from Eq.~(\ref{eq:C_E_first}) in order to obtain 
\begin{eqnarray}
\label{eq:C_E_firstandsecond_order_integrated_2}
 \deti \rho + \frac{1}{\Omega_d} \int  \partial_\beta \left(vu_\beta \rho \right)\de^{d-1}\uf
&-& \frac{\tau}{\Omega_d} \int \deti \partial_\alpha \left(v u_\alpha \rho \right)\de^{d-1}\uf\nonumber\\
&-& \frac{\tau}{\Omega_d} \int  \partial_\alpha \left( v u_\alpha \partial_\beta \left(v u_\beta \rho \right) \right) 
  \de^{d-1}\uf = 0 .
\end{eqnarray}
 The second and the third integral in the above expression are zero by symmetry
so that, using
\begin{equation}
\frac{\int u_\alpha u_\beta \de^{d-1}\uf}{\Omega_d} 
= \frac {\delta_{\alpha \beta}}{d} ,
\end{equation}
we are left with  
\begin{equation}
\label{eq:hydrodynamics2}
\deti \rho = \frac{\tau}{d}\nabla \cdot\left[v\nabla(\rho v)\right] .
\end{equation}

Eq.~(\ref{eq:hydrodynamics2}) is at the basis of the physics of 
self-trapping, and has some remarkable
consequences~\footnote{We note that the theory we have derived implicitly assumes that there is
no macroscopic polarisation of the self-propelled particle orientation.}.  
%We first observe that for constant running velocity $v$ we would obtain 
%the effective diffusion constant $D_{eff} = v^2 \tau / (d \Omega_d)$. THis results has to be compared with ...???
The important point here is that the running velocity can depend on the position, that is 
$v = v(\x)$. In steady state conditions, the distribution (obtained by setting
$\deti \rho=0$), is $\rho = 1/v(\x)$ -- hence bacteria, or self-propelled particles,
accumulate where they go slower. This may seem intuitive, but is an
important property which
has no counterpart in passive suspensions of Brownian particles. Indeed,  consider
a system of non-interacting Brownian particles, which have a position dependent
diffusivity, $D(\x)$. Because the system is passive, in equilibrium the probability
distribution (hence the density) needs to be proportional to the Boltzmann weight,
and this would give a uniform density in absence of any interaction.
%
%
%i.e. to $\exp(-V/k_BT)$, where $T$ is temperature, $k_B$ is the Boltzmann constant, 
%and $V$ is the potential. Because the particles are non-interacting [just 
%as our swimmer in Eq.~(\ref{eq:swimmerskinetic})], $V=0$ hence it follows 
%that the density has to be uniform. 
Mathematically, this is because the correct
coarse grained equation for the density of colloidal particles $\rho(\x)$ would be
  \begin{equation}
\label{eq:hydrodynamicspassive}
\deti \rho = \nabla \cdot\left[ D(\x) \nabla \rho\right], 
\end{equation}
i.e. the diffusion and the velocity enter 
in different positions with respect to the gradients!
As a result, in steady state the only possible solution is $\rho$ equal to
a constant, at variance with the case of the self-propelled particles.

It turns out that this seemingly technical difference leads to dramatic
differences in the physics of motile and Brownian systems. Consider for instance
the case of a concentrated suspension of active particles, where the swim speed decreases
for density, for example,  simply due to crowding. 
Let us consider for simplicity an exponential decay with density,
% $v(\rho)$
\begin{equation}\label{vrho}
v \sim {\eu^{-\lambda\rho/2}}v_0 , 
\end{equation} 
where $\lambda$ measures how steep is the decay of the swim speed with
$\rho$. Eq. (\ref{eq:hydrodynamics2}) in 1D then becomes:
\begin{equation}
\label{eq:hydrodynamics3}
\deti \rho =  \partial_x \left[\tau v^2\left(1-\frac{\lambda\rho}{2}\right)\partial_x\rho \right] = \partial_x \left[D_{eff}(\rho)\partial_x\rho\right] .
\end{equation}
The effective diffusion coefficient $D_{eff}$ is now a function of the density. More importantly, for $\rho$ high enough $D_{eff} < 0$, 
so the effective diffusivity becomes negative! Negative diffusion means that an infinitesimally small fluctuation leads to a divergent
increase in the density field. In other words, the system phase separates. 
A regularised version of 
Eq.~(\ref{eq:hydrodynamics3}) can be written down by introducing a fourth-order derivative, as follows,
\begin{equation}
\label{eq:hydrodynamics_regular}
\deti \rho =   \partial_x \left[D_{eff}(\rho)\partial_x\rho\right] - k \partial^4_x\rho
\end{equation}
The latter term is analogous to the surface tension between particle-rich and particle-depleted domains appearing in standard continuum theories for multiphase fluids~\cite{ChaikinLubensky}.

Physically, the phase separation described by Eq.~(\ref{eq:hydrodynamics_regular}) can be understood as due to the positive feedback mechanism described in the introduction.
%which works as follows. Imagine that, due to some small density fluctuation, there is a region where swimmers are more %%concentrated. In this 
%region, their swim speed will decrease (due to our choice of $v(\rho)$). Now, 
Due to the fact that the swimmers accumulate where they go slower,
the local density will increase leading to more decrease in velocity and eventually to phase separation.
In reality this happens for
a sufficiently steep decrease of swim speed with density, 
as predicted by Eqs.~(\ref{eq:hydrodynamics3},\ref{eq:hydrodynamics_regular}).
This mechanism, based on the appearance of an  instability threshold  for the diffusivity,
%described by Eq.~(\ref{eq:hydrodynamics3})
 has been named ``self-trapping'', and the phase separation itself is sometimes called activity-induced, or motility-induced,
phase separation~\cite{Cates12,Tailleur08}. 

Eq.~(\ref{eq:hydrodynamics2}) and Eq.~(\ref{eq:hydrodynamics_regular}) have been derived for run-and-tumble particles, such as bacteria;
  another important class of self-propelled colloids is that of ``active Brownian particles'' (henceforth ABPs). In ABPs, the direction of swimming only changes gradually, by rotational diffusion (as opposed as to ``instantaneous'' tumbles) -- indeed the term Brownian comes from the fact that the typical origin of the gradual change of swimming direction is rotational diffusion. Like run-and-tumble particles, ABPs also exhibit motility-induced phase separation (see Section 3, and, for instance, the theory and simulations in~\cite{Tailleur08,valeriani2011colloids}). It is important to note here that the same theory leading to Eq.~(\ref{eq:hydrodynamics2}) and Eq.~(\ref{eq:hydrodynamics_regular}) can be modified in a rather straightforward way to apply to ABPs, essentially by substituting $1/\tau$ with $(d-1)D_R$~\cite{Tailleur08,Tailleur10} ($D_R$ is the rotational diffusion coefficient).

%%%%%%%%%%%%%%% Continuum model %%%%%%%%%%%%%%%%%%%%%%%%%%%

\subsection{Relation with Cahn-Hilliard equation, noise and interfacial terms}

The approach adopted so far neglects hydrodynamic noise, and it further lacks a systematic derivation of the $- k \partial^4_x\rho$ term, or indeed of any interfacial terms, which should be of fourth-order in gradients.

Here we briefly sketch the theory, initially derived in Ref.~\cite{Tailleur08,Tailleur10} and afterwards 
generalized in Ref.~\cite{Stenhammar13,Stenhammar15}, in which these aspects have been included.

First, we note that Eq.~(\ref{eq:hydrodynamics2}) can also be written formally as a Cahn-Hilliard equation, 
\begin{equation}\label{continuum_eqn}
\partial_t\rho=-\nabla\cdot\left[-M(\rho)\nabla\mu\right].
\end{equation}
In Eq.~(\ref{continuum_eqn}), the chemical potential $\mu$ can be obtained from the functional differentiation of an effective free energy ${\cal F}$ with respect to $\rho$, with ${\cal F}=\int f_0 d{\bf r}$, and with $f_0=\rho(\ln\rho-1)+\int_0^{\rho}\ln[v(\rho')]d\rho'$ (this was first noted in~\cite{Tailleur08}). The term $M(\rho)$ is an effective mobility, equal to $\frac{v^2\rho\tau}{d}$ for run-and-tumble particles (recall from Section 2.1 that $\tau$ is the run duration, equivalently the inverse tumbling rate). An explicit calculation of the chemical potential as $\mu=\delta F/\delta \rho$ and a bit of algebra show that Eq.~(\ref{continuum_eqn}) is the same as Eq.~(\ref{eq:hydrodynamics2}).

Within the effective Cahn-Hilliard description in Eq.~(\ref{continuum_eqn}), the role of fluctuations can be included by added a conserved noise as follows (see~\cite{Tailleur08} for details),
\begin{equation}\label{continuum_eqn2}
\partial_t\rho=-\nabla\cdot\left[-M(\rho)\nabla\mu+\sqrt{2M(\rho)}{\bf \Lambda}\right].
\end{equation}
On the right hand side of the equation ${\bf \Lambda}$ is a noise vector with components $\langle\Lambda_i({\bf r},t)\Lambda_j({\bf r'},t')\rangle=\delta_{ij}\delta({\bf r}-{\bf r}')\delta(t-t')$.

Importantly, a full analysis of the phase separation kinetics requires the presence of interface-like terms to stabilise domain walls between the two phases. This is achieved by assuming that $v$ is not strictly local but $v=v(\hat{\rho})$, with $\hat{\rho}({\bf r})=\rho+\gamma^2\nabla^2\rho$.  This is the leading order correction allowed by rotational invariance~\cite{Stenhammar13}. Here $\gamma$ is a smoothing length proportional to the run length for run-and-tumble particles, or to the persistence length of ABPs: in formulas one expects $\gamma(\rho)=\gamma_0\tau_rv(\rho)$, where $\gamma_0$ is of order unity and $\tau_r$ is the orientational relaxation time (equal to either $\tau$ or to $D_{R}^{-1}$).  
This assumption relies on the fact that a single ABP samples the density $\rho$ over a length scale much greater than the interparticle spacing and is
proportional to the persistence length $l(\rho)=\tau_rv(\rho)/(d-1)$.

By using the nonlocal form of $v(\rho)=v(\hat{\rho})$ just advocated, it can be shown that the chemical potential including interfacial terms may be written in the following form~\cite{Stenhammar13}
\begin{equation}\label{noneqmu}
\mu=\ln\rho+\ln v(\rho)-\kappa(\rho)\nabla^2\rho+O(\nabla^4\rho),
\end{equation}
where $\kappa(\rho)=-\left[\frac{\gamma_0\tau_r}{(d-1)}\right]^2v(\rho)\frac{dv}{d\rho}$. It is important to note that the interfacial term breaks the mapping to a ``standard'' Cahn-Hilliard equation, in that there exist no free energy functional of which Eq.~(\ref{noneqmu}) is a functional derivative. Equivalently, the interfacial term breaks detailed balance. 

Finally, we note that in Ref.~\cite{Stenhammar13} a repulsive term of the form $f_{rep}=\omega\Theta(\rho-\rho_l)(\rho-\rho_l)^4$ is summed to $f_0$  to model excluded volume interactions, and to prevent the existence of a phase of infinite density. The repulsive free energy term is phenomenological, and it consists of constant $\omega$ multiplied by the Heaviside step function $\Theta$; there is also a density threshold $\rho_l$ at which the repulsion is switched on.

To make calculations in practice with the continuum model in Eq.~(\ref{continuum_eqn2}), an explicit expression for both $v(\rho)$ and $M(\rho)$ is needed. These require some information from microscopic theories, or simulations, which we will discuss next. Suffice here to say that, as shown in Ref.~\cite{Stenhammar13}, an appropriate choice at low and intermediate densities is the following:
\begin{equation}\label{vrhoABP}
v(\rho)=v_0(1-v_0\sigma_s\tau_c\rho),
\end{equation}
for $v(\rho)$, and
\begin{equation}
M(\rho)=\frac{v^2(\rho)\rho\tau_r}{d(d-1)}=D_0 \rho (1-v_0\sigma_s\tau_c\rho)^2,
\end{equation}
for $M(\rho)$. In the two equations above, $\sigma_s$ is a scattering cross-section. This quantity is obtained by assuming that a single ABP trajectory consists of straight runs with speed $v_0$ punctuated by collisions lasting for a time $\tau_c$ in which the particle arrests its motion. (We note that Eq.(\ref{vrhoABP}) is also consistent with the limit of Eq.(\ref{vrho}) for $\lambda\to 0$.)

\begin{figure}[h]
\includegraphics[width=1.0\textwidth]{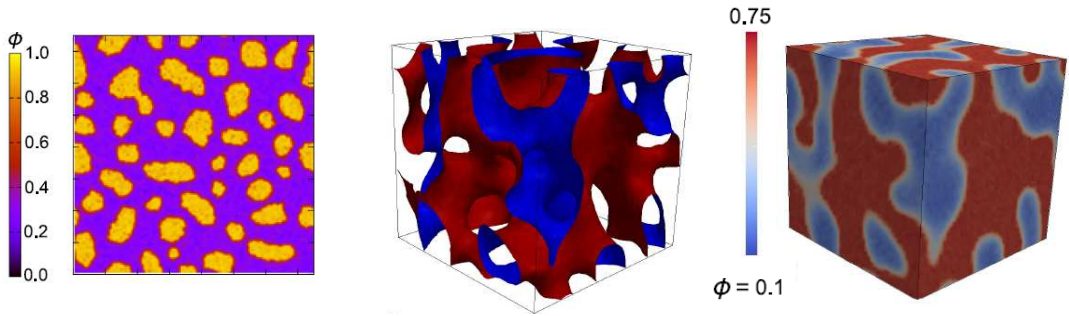}
\caption{In this figure we report snapshots for the local packing fraction $\phi=\rho/\rho_0$, with $\rho_0=(v_o\sigma_s\tau_c)^{-1}$,
obtained by numerically solving Eq.~(\ref{continuum_eqn2}) in 2D (left panel) and in 3D (middle and right panel), both at a value of the overall packing
fraction $\phi_0=0.5$.
The right  panel shows a  2D projection of the corresponding 3D density field, whose isosurfaces are shown in the middle panel.
More details about the mapping between $\phi$ and $\rho$ are described in Ref.~\cite{Stenhammar13}.}
\label{fig:continuum_model}
\end{figure}

In Fig.~\ref{fig:continuum_model} we show intermediate-time snapshots of the concentration $\phi=\rho/\rho_0$ (where $\rho_0=(v_o\sigma_s\tau_c)^{-1}$)
obtained by numerically solving Eq.~(\ref{continuum_eqn2}) in 2D (left panel) and in 3D (middle and right panel) both at overall packing fraction
$\phi_0=0.5$~\cite{Stenhammar13,Stenhammar15}. Isolated domains of the dense phase in 2D (in yellow), clearly distinguished from the diluite phase (in blue), 
represent regions where particles tend to accumulate. Especially in 3D, the domain topology resembles one which is expected of phase separation in passive gas-liquid systems, or binary fluids. 

It is quite a remarkable and striking feature that the model in Eq.~(\ref{continuum_eqn2}), although almost fully built within the framework of equilibrium statistical thermodynamics~\footnote{The weak transgression from equilibrium and detailed balance is the interfacial term, which has been extensively discussed in Ref.\cite{Stenhammar14}. This has been shown to have no impact on the coarsening dynamics (see Fig. 5).}, captures very well the existence and dynamics of motility-induced phase separation, which is a far from equilibrium phenomenon. (Indeed, as we shall see more in detail in Section 3, Eq.~(\ref{continuum_eqn2}) also reproduces well the dynamics observed in direct particle simulations of concentrated suspensions of active Brownian particles, with an enourmously reduced computational cost~\cite{Stenhammar15}.)

\section{Motility-induced phase separation in self-propelled particle systems}
\label{sec:selfpropelled_particles}

\begin{figure}[ht]
\includegraphics[width=0.9\textwidth]{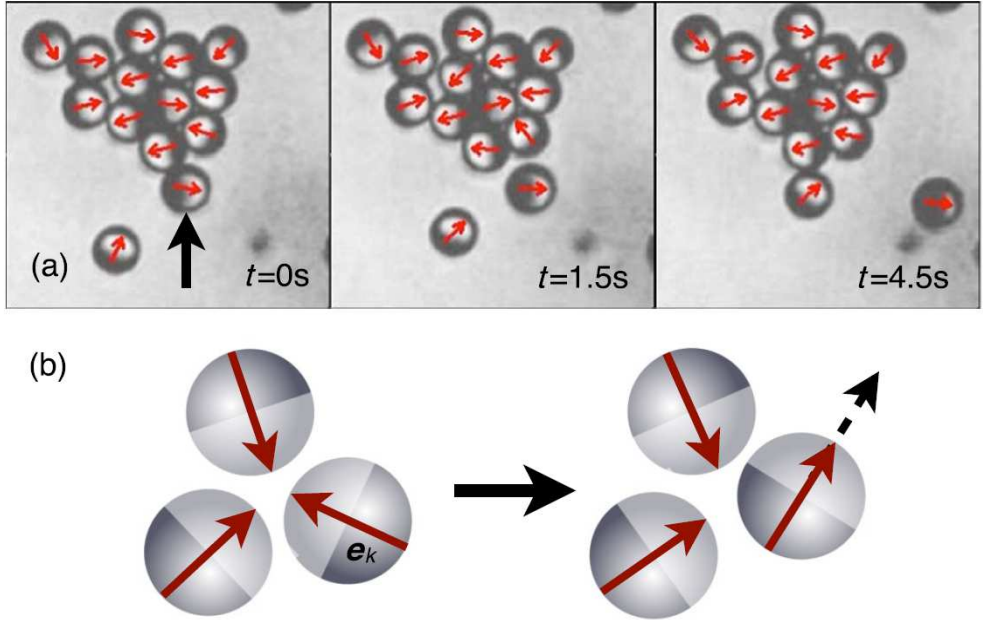}
\caption{(a) An example of the dynamics of a cluster of active colloids. The particle marked by the black arrow  (left snapshot)
leaves the cluster and is substituted by another particle approaching (right snapshots). Red arrows denote the
orientation of the caps. (b) The mechanism used by a particle for self-trapping is shown.
After a collision, a particle becomes free when  its orientation changes due to the rotational diffusion.
This figure is taken from Ref.~\cite{Buttinoni13b} 
where large enough Janus particles, allowing to visually resolve 
the single polarizations, are used.}
\label{fig:prl_buttinoni}
\end{figure}

In practice, both in simulations and in experiments, motility-induced phase separation has been observed in active brownian particles, rather than in run-and-tumble bacteria (recall that in ABPs rotational diffusion replaces tumbling, see Section 2). 
ABPs can be realised in practice, for instance, as synthetic (bi-)metallised colloids where some chemical reaction creates a self-phoretic local chemical motor\footnote{Self-phoretic motors are particles which move due to the gradients of a chemical which they produce themselves.}~\cite{Walther,Paxton04,Paxton06,Hong06,Howse07,Golestanian07,Palacci10}.  
An example of a motility-induced clusters observed experimentally in such systems is shown in Fig.~\ref{fig:prl_buttinoni}. 
It is also relatively straightforward to set up a Brownian simulation of ABPs, and because this is the approach which has been most commonly employed in the study of MIPS, we focus in this Section on numerical work, reviewing first the model, then the phase diagram for MIPS, and later on the dynamics of MIPS in ABPs (Sections 3.1-3.2 describe simulations for spherical particles, 3.3 and 3.4 describe more general models for rodlike particles, or self-propelled particles with an alignment interaction).

In the simplest theoretical framework leading to MIPS, $N$ spherical ABPs interact with each other  only by excluded volume effects and are subject to a propulsive  or active force continuously acting  along a fixed polar  axis for each particle~\footnote{In a system with momentum conservation, as discussed later, the total force on a neutrally buoyant swimmer should instead be zero. However Brownian dynamics theories, e.g. simulations, neglect fluid-mediated interactions so the only way to propel a particle is to apply a force along its direction.}. This  force is typically constant in modulus  but its direction changes with time due to rotational diffusion.
A minimal theoretical model with these characteristics is the following 
\begin{equation}
\dot {{\mathbf r}}_i = \hat {\boldsymbol \nu}_i v_0   + \mu \sum_{j \ne i} {\mathbf F}_{ij} + {\boldsymbol \eta}^{\rm T}_i \quad ,
\label{eq:activecolloids1}
\end{equation}
\begin{equation}
\dot \theta_i =  {\mathbf \eta}_i \quad .
\label{eq:activecolloids2}
\end{equation}
Here ${\mathbf r}_i$ are the coordinates of the center of mass of the particle $i$ ($i = 1,...,N$) and $\mu$ is the mobility, while 
the polarity axis of each particle is defined by $ \hat {\boldsymbol \nu}_i = (\cos \theta_i, \sin \theta_i)$,
where the angle  $\theta_i$ is measured with respect to some fixed laboratory axis. Furthermore, $v_0$ is the fixed self-propulsion velocity and
 $ {\mathbf F}_{ij}$ is the force due to the particle $j$ acting on the particle $i$ that takes into account excluded-volume steric effects.
${\boldsymbol \eta}^{\rm T}_i(t)$ and $\eta_i(t)$ are Gaussian, white noises with zero mean and correlations 
$<\eta^{\rm T}_{i \alpha}(t) \eta^{\rm T}_{j \beta}(t')  > = 2 D \delta_{ij} \delta_{\alpha \beta} \delta(t- t')$
and $ < \eta_i(t) \eta_j(t')> = 2 D_R \delta_{ij} \delta(t-t')$, where $D = k_B T \mu $ and $D_R$ respectively denote the brownian translational diffusion constant ($T$ is the temperature of the bath) and the single particle rotational diffusion constant $D_R = 3 D /\sigma_c^2$, with $\sigma_c$ the diameter of the colloid.   

The occurrence of  phase separation in a realization of the above model was first shown in~\cite{Fily12}, and later on confirmed and analysed in several works (which are reviewed later on). A nice feature of particle-based studies of MIPS, whether experimental or numerical, is that these allow a clear identification of the microscopic mechanism leading to the initial formation of clusters of self-propelled particles.
In particular, one observes that when two or more  particles collide head-on they are blocked due to the persistence of their orientations (see the experimental pictures in Fig.~\ref{fig:prl_buttinoni}). 
A particle can escape from the cluster if, due to rotational diffusion, the 
direction of the polar axis changes so that the active force pushes away the 
particle from the cluster.
This microscopic view is complementary to the macroscopic theory of MIPS, presented in Section 2.

Relevant adimensional numbers in terms of which the above model can be analyzed are the following. The  surface fraction is given by
\begin{equation}
\phi_0=N \ \frac{S_{\rm c}}{S} 
\;  
\end{equation}
with $S_{\rm c}$ the area occupied by an individual colloid and  $S$ the total area of the box where the particles  move. 
 Then we introduce the P\'eclet number, ${\rm Pe}$,  which  is a dimensionless ratio between the  
advective transport rate and the diffusive transport rate.
For particle flow one defines it as 
$ {\rm Pe} = Lv/{D} $,
with $L$ a typical length, $v$ a typical velocity, and $D$ a typical diffusion constant. We choose $L \to \sigma_{\rm c}$, $v \to v_0$, and 
$D \to D = k_BT \mu)$. Then,
\begin{equation}
{\rm Pe} = \frac{\sigma_{\rm c} v_0} {k_B T \mu}
\; . 
\label{eq:Peclet}
\end{equation}
While other definitions of the  P\'eclet number based on rotational diffusion are possible, we use here 
the same definition given in the references discussed  in this section.

\subsection{Phase diagram}

\begin{figure}[ht]
\begin{centering}
\includegraphics[scale=1.2]{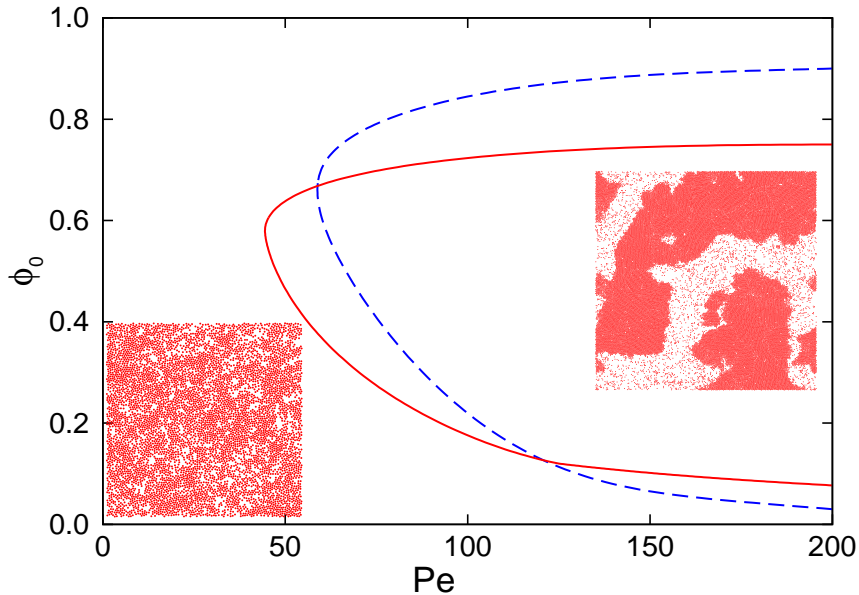}
\end{centering}
\caption{(Color online.) {Phase diagram of a system
of spherical ABPs (red continuous curve), and of dumbbells (blue dashed curve).
Inside the curve, at high P\'eclet numbers, the system undergoes
phase separation into two phases characterized by  two different densities.
Here, typically, large  and stable clusters are observed, as shown in
the snapshot for spherical ABPs on the right. In the case of
dumbbells, these are frozen due to steric interactions and point
preferentially
towards  the center of the cluster, while active colloids can still
freely rotate inside the cluster. For small P\'eclet numbers
 the system does not show the formation of such large and stable
clusters; a typical
snapshot of the system in this uniform phase is also shown, again for
spherical ABPs.
The location of the transition line is based on the results of
simulations fixing $v_0=0.1$ in both cases and varying T (see
\cite{Suma13, Suma14} for further details);
the right snapshot is taken at $T=0.01$, the one on the left at
$T=0.05$.
Spherical ABPs  follow an equation of motion with the same
parametrization 
%(self-propulsion velocity, mobility, temperature, etc)
 of the center of mass of  dumbbells.}
}
\label{fig:phasediagram} 
\end{figure}

To begin our review of particle-based simulations of MIPS using the model described above, we discuss determinations of the phase diagram.
From the considerations made above, and in Section 2, we expect that there should be a phase transition (MIPS) at low temperature or large self-propulsion (i.e. at large P\'eclet number) when the density of the system is sufficiently high. The  phase diagram of a system of self-propelled disks is shown in Fig.~\ref{fig:phasediagram}. 
Here, the physical meaning of the red curve is similar to that of the coexistence curve in fluids with liquid-vapour transition.  Outside the coexisting line, the system can only be in a single phase, while for P\'eclet numbers higher than a critical value and densities inside the coexistence curve, it phase separates into a gas phase and a particle-rich phase. An approximate analytical estimate for the phase diagram can be found in Ref.~\cite{Baskaran12} -- while the theory in Section 2 predicts phase separation, it disregards thermal diffusion so strictly works only for very large (actually infinite) P\'eclet number.

An interesting quantitative prediction of the phase diagram in Fig.~\ref{fig:phasediagram} is that the critical point occurs for a value of the P\'eclet around 50, which can quite
readily be realised with e.g. bacterial swimmers (the challenge though is
to concentrate them up to $\phi_0 \sim 0.6$, such  
that MIPS is predicted to occur in Fig.~\ref{fig:phasediagram}).

To determine numerically the coexistence curve one may, for instance,
 consider the probability distribution function $P$ of the local density $\phi$. 
This can be obtained dividing the full system  in  cells  with linear size 
larger (typically 10 times) than that of each particle and calculating 
the density in each cell. The size of the cells has to be  much smaller than
the linear size of the full sample and big enough to sample $\phi$ correctly.
Then, letting the system evolve from an initial  homogeneous configuration  
at a given total surface fraction, one can measure the distributions of 
local densities, $P(\phi)$, in steady state. 

\begin{figure}[ht]
\begin{center}
  \begin{tabular}{cc}
    \includegraphics[scale=0.67]{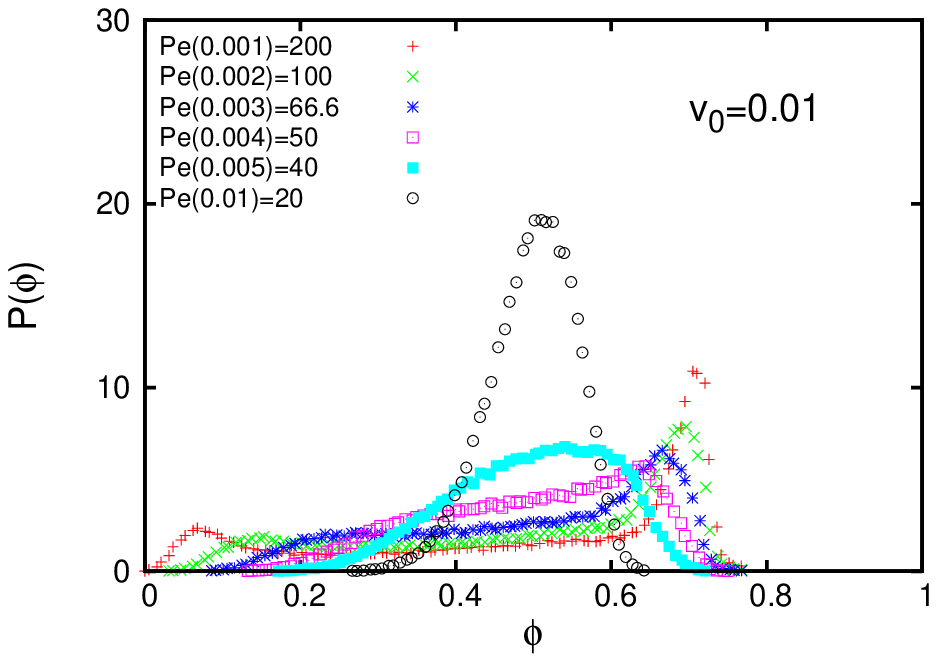} 
     \includegraphics[scale=0.67]{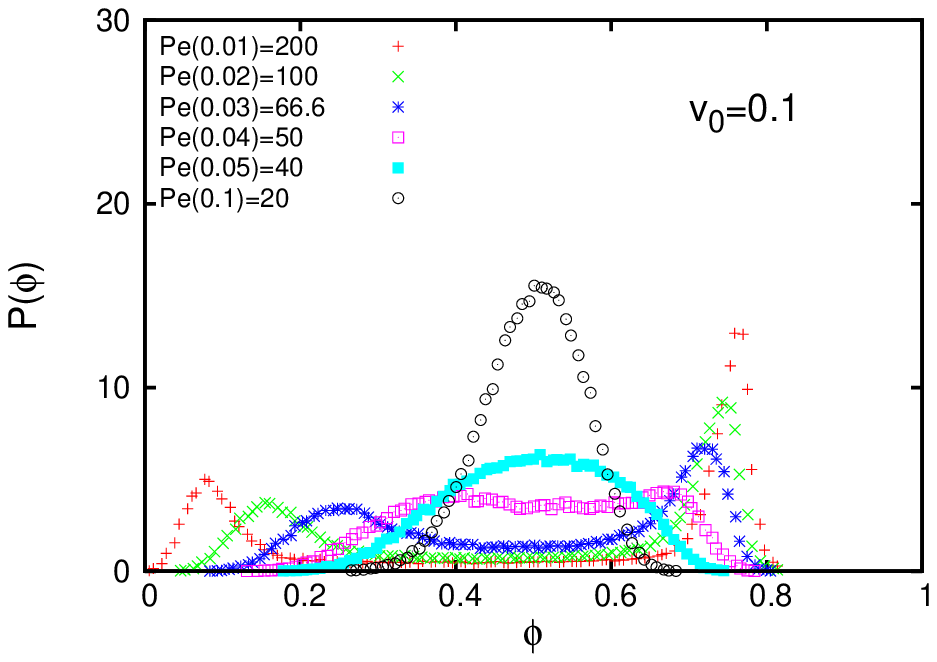}
    \end{tabular}	
\caption{Local density distributions for spherical ABPs at P\'eclet numbers given in the keys in
the form Pe$(T)$
for different temperatures,  $v_0=0.01, 0.1$, $\sigma_c=1$,  and $\mu=0.05$.
The global density of the system is $\phi_0=0.5$. }		
\label{fig:density_distributions} 
\end{center}
\end{figure}

An example of the analysis of these distribution functions, for a system of 
self-propelled disks is shown in Fig.~\ref{fig:density_distributions} 
(the corresponding parameters are given in the caption). 
In each  panel of this figure one observes that at low P\'eclet number the 
distribution is characterized by a single peak approximatively corresponding 
to the value ot the total density while at high P\'eclet the distribution is
characterized by the presence of two peaks which can be taken as the density 
values for the two coexisting phases. This can be shown to be a consistent 
procedure, e.g. by checking that starting from a different total surface 
fraction  the values of the two peaks remain the same~\cite{Suma14,Suma15}. 

\vspace{0.75cm}

\subsection{Kinetics of phase separation}

We now turn to the review of the studies of the kinetics of MIPS (these are less numerous in the literature). In analogy with the study of domain growth and coarsening in gas-liquid systems and binary fluids, the idea here is to measure the size of the phase separating domains (the particle-rich ones) $L(t)$, for instance by taking appropriate moments of the structure factor~\cite{ChaikinLubensky}. 
The size of these domains then increase in time, and the analogy with Cahn-Hilliard models suggests that such a dynamics follows a power law,
\begin{equation}
L(t) \sim t^{\alpha},
\end{equation}
where $\alpha$ is the exponent which measures how quickly coarsening proceeds. It is useful to recall here that the passive counterpart of the ABP model we are considering is a gas-liquid system in the absence of solvent-mediated hydrodynamic interactions -- under these conditions one expects $\alpha=1/3$~\cite{ChaikinLubensky}.

In their ABP simulations, Redner et al.~\cite{Redner13}, measured a growth of the cluster mean-size compatible with an exponent $\alpha = 0.272$. 
Stenhammar et al.~\cite{Stenhammar15}, in a variant of the model of  (\ref{eq:activecolloids1},\ref{eq:activecolloids2}),
performing  more extensive simulations with $5 \times 10^5$ particles,  found $\alpha = 0.28 $ in  $2D$ systems.
The same authors also performed  3D simulations with  $4  \times 10^7$ particles  obtaining in this case $\alpha= 0.34 $.
 
The previously mentioned  power-law regimes were  found after 
an initial transient with faster growth. It was observed in~\cite{Stenhammar15}  that the cross-over to the final regime 
occurs when $L(t)$ exceeded a persistence length related to the characteristic  balistic displacement  of a single particle.
This corresponds to the persistence length mentioned in Sect.2.2 and  can be evaluated  as $l= v_0/(D_R(d-1))$.  Since this quantity  is proportional to the P\'eclet number
(it depends on the ratio $v_0/T$) and phase separation occurs  with ${\rm Pe} > {\rm Pe}_c$, it follows
that  very big systems are needed in order to have a large enough extension of the late time regime 
with  reliable measurements. For example, in the simulations of~\cite{Stenhammar15}, 
Pe=300 implies  $ l = 50 \sigma$  so that  a linear size $L_{\rm box} =350 \sigma$  
was considered leading to $4 \times  10^7$ particles in order to have $\phi_0=0.5$.  

In general, the results in both 2D and 3D are interpreted as consistent with the $\alpha=1/3$ exponent, which is expected for very long times
(such times are reached in 3D, but not in 2D due to the reasons just discussed).
A natural question is whether the run-and-tumble particles previously discussed behave in the same way. In this area,
the largest simulation to date is that in Ref.~\cite{Thompson10}, which uses a lattice model of run-and-tumble particles and
demonstrates that the ``diffusive'' exponent $\alpha=1/3$ is found in simulations of MIPS in 2D. 

\begin{figure}[ht]
\includegraphics[width=1.0\textwidth]{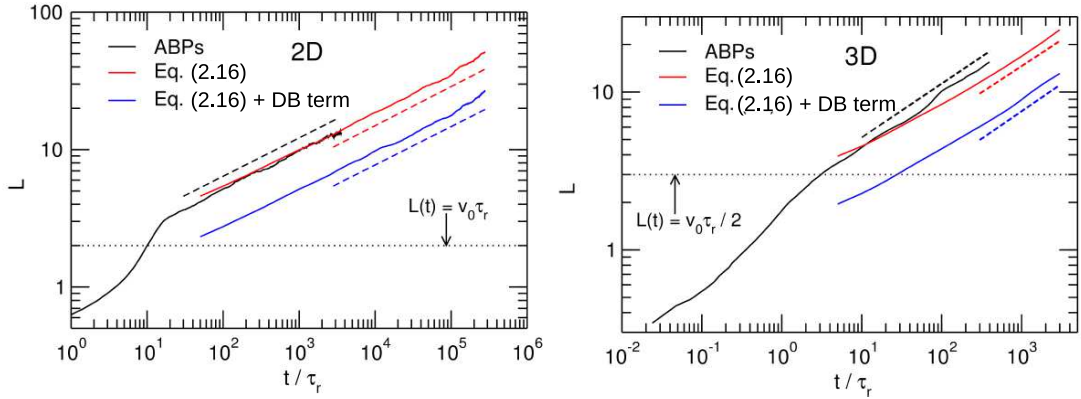}
\caption{In this figure the length-scale $L(t)$ for the 2D (left snapshot) and the 3D (right snapshot) systems are shown. Both of these are obtained for packing-fraction $\phi_0=0.5$. 
The black curves (top set of curves, starting from the leftmost point in the figures) refer to $L(t)$ as found by particle based simulations (ABPs); the red curves (middle set of curves in the figures) refer to continuum model simulations according to Eq.(2.16); the blue curves (bottom set of curves in the figures) refer to $L(t)$ for the case in which the detailed balance is restored (see~\cite{Stenhammar14}).
The dashed lines gives the corresponding fitted exponents, which are: $\alpha_{2D}=0.279$ and $\alpha_{3D}=0.341$ for ABPs, $\alpha_{2D}=0.287$, $\alpha_{3D}=0.333$ for the continuum model with detailed balance violation.
 $\tau_r$ is the orientational relaxation time defined in Sect. 2.2. The dotted lines represent the persistence length which separate the region of superdiffusive behaviour ($\alpha>1/3$) at short times from the region of diffusive behaviour ($\alpha\simeq 1/3$).
Figure is taken from Ref.~\cite{Stenhammar15}, with permission.}
\label{fig:softmatter_joak}
\end{figure}

\subsection{MIPS with non-spherical particles: active dumbbells}

Until now, we have focussed our review on the case of spherical self-propelled particles. In active matter physics, however a spherical shape as assumed in the previous section, is the exception rather than the rule. Swimmers, whether synthetic  or natural, are often rod-like or elongated: this is true both of most bacteria, but also for chemically propelled nanorods~\cite{Fletcher09,Ramaswamy10,Vicsek12,gompper14}. It is therefore of interest to analyse how MIPS is affected when the self-propelled particles have an elongated shape, i.e. they are rods, or dumbbells. In particular we will consider in this Section the case of active dumbbells.
 
In Ref.~\cite{valeriani2011colloids} a model of active dumbbells was introduced to describe the diffusive experimental behaviour of a bacterial bath coupled to colloidal tracers. Other simulations of rodlike active particles were presented in~\cite{Peruani12prl,Baskaran12,Wensink12}: all these cases demonstrate that the physics of anisotropic active particle is richer than that of their isometric counterparts, for instance due to the possible onset of local orientational (nematic) ordering. To the best of our knowledge, the first analysis of the effect of shape anisotropy on MIPS was performed in~\cite{Suma14}.
In that work, the phase behaviour of active dumbbells of variable overall density and P\'eclet number was studied by means of Brownian dynamics simulations.
% (Ref.~\cite{Suma14b} provides a complementary study of the same model by using the LAMMPS Molecular Dynamics package~\cite{Plimpton}). 
The model of active dumbbells is a simple generalisation of those of ABPs; however, for dumbbells the rotational diffusion is no longer
explicitely described as in Eq.~(\ref{eq:activecolloids2}), rather it effectively results from the net balance of forces on the two disks of each dumbbell (a detailed description of the model is given
 in~\cite{Suma14b}). 
Rotational and translational diffusion properties of active dumbbells have been studied in ~\cite{Suma14b,Suma15,Suma15b}.

The elongated shape of dumbbells leads to some important differences with
the case of spherical active particles.
To begin with, phase separation is overall favoured in the case of dumbbells~\cite{Suma14}. For instance, in Fig.~\ref{fig:phasediagram} the coexistence curve for a dumbbell system can be compared with that of colloids; while the curves cross close to the critical point, for sufficiently large Pe the density range whether MIPS can be observed is larger for dumbbells. 

\begin{figure}[!ht]
\begin{center}
\includegraphics[width=0.8\columnwidth]{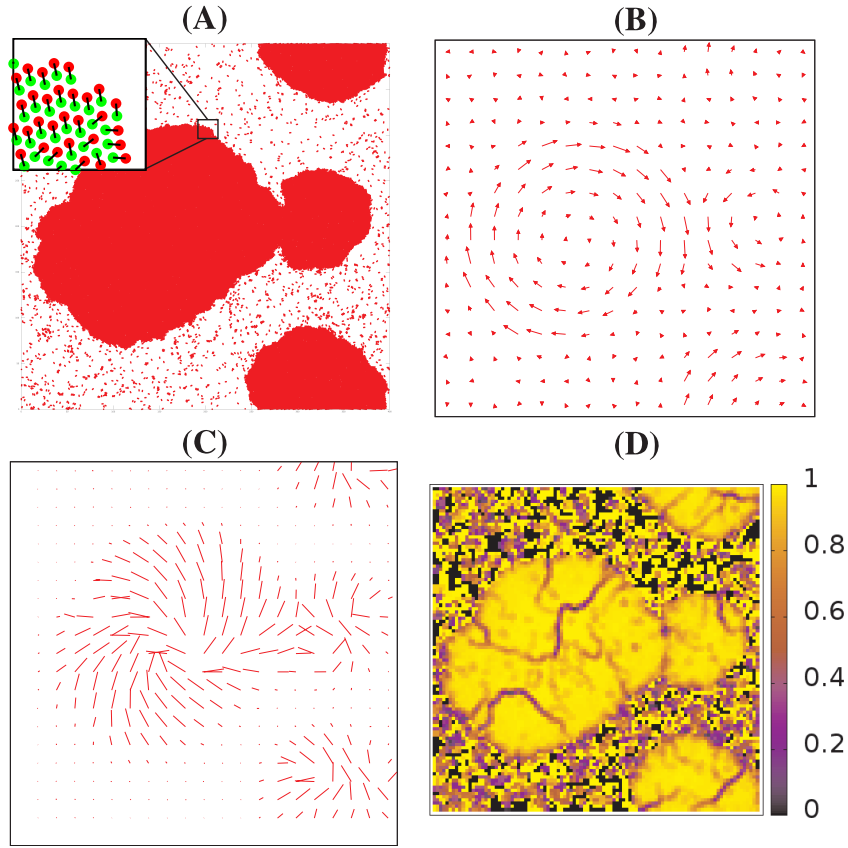}
\end{center}
\caption{(A) Snapshot of a phase-separated active dumbbell fluid with three
clusters (two of which are touching). The clusters, from the smallest to the largest one, contain
about 4000, 12000, and 25000 dumbbells, respectively.
 The inset shows a detail
of the dumbbell configuration; red and green beads indicate the tail and head
{of each dumbbell and are connected by a line.}
(B) Coarse grained velocity field corresponding
to the snapshot in (A), obtained by local averages on a square mesh. (C) Coarse-grained orientation profile
of the dumbbell fluid (${\mathbf P}$), obtained by averaging as in (B), again 
corresponding to the configuration in (A). 
(D) Hexatic order parameter (computed as in Ref.~\cite{Redner13})
corresponding to the configuration in (A) 
(defects, or grain boundaries, 
betweeen ordered regions are observable).
Figure is taken from Ref.~\cite{Suma14}. 
}
\label{fig:rotating_clusters}
\end{figure}

However, by far the most striking feature about the  dumbbell aggregates is
that they exhibit coherent and long-lived vortex-like rotation (see Fig.~\ref{fig:rotating_clusters}).
This coherent rotation is absent in the motility-induced clusters formed with spherical particles. On the other hand, we also observe that 
the clusters observed in  the system of active rods of Ref.~\cite{Peruani06,yang2010swarm,Baskaran12}, where
the length-to-width  aspect ratio is larger compared with that of our dumbbells,
show a translational motion but little detectable rotation.

What drives the self-organisation of rotating aggregates in systems of dumbbells? And why are they absent for spherical swimmers?
A first explanation for the formation of such structures 
comes from considering torque balance
on the droplets which form during MIPS. 
Within the Brownian dynamics simulations performed in~\cite{Suma14}, the 
forces exerted by each dumbbell add a local torque
which is balanced by the drag on the cluster. 
While also spherical self-propelled particles would exert a local
torque, this can only lead to rotational diffusion and not to
sustained rotations, because the local swimming direction of
each active particle can rotate freely within the 
cluster~\cite{Redner13}.
This is not the case for the dumbbell aggregates, where steric 
interactions essentially quench the polarisation of each of the dumbbells
in the rotating droplet (see Fig.~\ref{fig:rotating_clusters}). 

The overall coherent rotation of the cluster in 
Fig.~\ref{fig:rotating_clusters} is apparent from a coarse grained plot of the
dumbbell momentum field $\rho {\mathbf v}$ (see 
Fig.~\ref{fig:rotating_clusters} B), 
where $\rho$ and ${\mathbf v}$ are the local density and velocity 
field of the dumbbells at a given point in space and instant in time.
It can be seen that the topology of $\rho {\mathbf v}$  is that of a vortex.
Inspection of the local ordering within the high density rotating clusters
of dumbbells also demonstrates that there is both nematic and hexatic
order~\cite{Suma14}. The local director profile, measuring 
the average orientation of the dumbbells, ${\mathbf P}$, 
shows a characteristic spiral pattern (Fig.~\ref{fig:rotating_clusters}C).
A distinguishing aspect of the  rotating aggregates in dumbbell systems  
is therefore that the velocity and orientational patterns are distinct: 
the former is a vortex, the latter a spiral. 
A minimal mean field framework to understand this fact was proposed in
Ref.~\cite{Suma14}: a conclusion from this theory is that a requirement
in order for polarisation and velocity field not to be parallel to 
each other is that the theory should include a non-trivial 
momentum balance between self-propulsion and steric repulsion. 

Computer simulations suggest that the velocity of the rotating clusters,
$\Omega$, should depend on their radius, $R$. In particular, the Brownian
dynamics simulations of Ref.~\cite{Suma14} measure a scaling of
$\Omega$ with $1/R$. This scaling can be understood by means of
dimensional analysis as follows. First, we note that the torque on the cluster must be proportional to 
$R^3 F_a$,  where $F_a$  is the magnitude of the active force acting on each dumbbell, as the individual dumbbell torque is $\sim RF_a$,
and  given that the polarisation within the aggregate is coherent (interestingly, other
arrangements, e.g. random, would lead to different scaling). 
Second, the rotational friction scales as $R^4$~\cite{Suma14}. Balancing the two contributions
leads to the observed $1/R$ scaling.
It is interesting to highlight here that rotating clusters have been found in
experiments with mixtures of bacteria and polymers, where bacteria feel an
effective attractive interaction due to depletion attractions arising from
the presence of the polymer~\cite{Schwarz-Linek12}. In the experiments,
again a $1/R$ behaviour of the angular velocity of the clusters is observed, however the
details of the balance leading to this scaling are different. In Ref.~\cite{Schwarz-Linek12},
bacteria aggregate in 3D, the torque comes from bacteria at the surface of the cluster, and the drag on the cluster
is Stokesian and $\sim R^3$. Furthermore, the rotating clusters found in Fig.~\ref{fig:rotating_clusters}
appear due to MIPS, whereas those in~\cite{Schwarz-Linek12} due to an additional interparticle attraction
(the depletion attraction).

While the work in Ref.~\cite{Suma14} covers the effect of shape anisotropy on the MIPS phase diagram and cluster morphology, an
interesting open question is to what extent and how shape may affect the kinetics of phase separation, e.g. the coarsening and
growth laws of the domains. 

Finally, it can be observed that  active dumbbells 
 differ from spherical ABPs because their shape introduces 
an additional hydrodynamic variable, the local polarisation ${\bf P}$. In this sense, they 
 can be viewed as a particular example of a more general
class of self-propelled particles, which interact with each other through an alignment interaction. 
In the case of dumbbells, this interaction comes from steric effects alone, but in other cases
it may have different origins: for instance bacterial swimmers may interact due to hydrodynamic
interactions~\cite{Ramaswamy10}. 
 In the best-known model for active particles, introduced by 
Vicsek et al. in 1995~\cite{vicsek},  an ensemble 
of self-propelled particles  interact solely via an alignment interaction whose origin
is unspecified. 
%The Vicsek  model is defined by a set of rules, according to which particles
%move along the direction of their instantaneous velocity, and update their velocity to align it
%to that of their neighbours (defined for instance as the particles within a radius $R_0$ from
%the particle under consideration), adding some noise.
% $\eta$. 
The Vicsek model exhibits a 
transition from an isotropic phase 
%at large noise 
to a flocking state
%, at low noise, where
with active particles moving coherently~\cite{Chate04,Baglietto08,Chate08,Ginelli10,Solon15}.
A review on different dynamical features in  the Vicsek model is beyond the scope of our current work 
and the interested reader can see e.g. Ref.~\cite{Chate08ejp}.
A generalization of the Vicsek model in which effective steric interactions among active particles are incorporated
is studied in  
%MIPS for
%active particles with explicit alignment interaction has been studied  in 
Ref.~\cite{Farrell12}.

%\begin{figure}[ht]
%\includegraphics[width=1.0\textwidth]{diagramma_di_fase/snap_dumbbells.eps}
%\caption{}
%\label{fig:domain_dumbbells}
%\end{figure}

%\begin{figure}[ht]
%\includegraphics[width=0.7\textwidth]{diagramma_di_fase/radius_6x8.epsi}
%\caption{}
%\label{fig:radius}
%\end{figure}

\section{Other aspects of motility-induced phase separations}

\subsection{MIPS with reproducing particles: arrest of coarsening}

In Sections 2 and 3, we saw that the physics of motility-induced phase separation is akin to that of phase separation in a gas-liquid system, or in a binary mixture. This is because the clusters, which form when active particles bump into one another, coarsen, e.g. by Ostwald ripening -- i.e. smaller aggregates evaporate to increase the size of larger clusters. In some cases, however, MIPS can also be arrested, so that clusters coarsen until a certain equilibrium size is reached, and no more. Here we consider one interesting example in which this happens: a population of reproducing run-and-tumble self-propelled particles whose velocity decreases with density~\cite{Cates10}.

A simple way to describe self-replication in bacterial colonies is to do so via a logistic growth law. This can be done naturally within our hydrodynamic framework, and leads to the following mean-field equation for the bacterial density,
\begin{equation}\label{logistic}
\deti \rho =  \partial_x \left[D_{eff}(\rho)\partial_x\rho\right] - k \partial^4_x\rho +\alpha\rho\left(1-\frac{\rho}{\rho_0}\right),
\end{equation}
where $\alpha$ is the growth rate (in $s^{-1}$), and $\rho_0$ is an equilibrium density which the bacteria tend to in the absence of other interactions. The value of $\alpha$ may vary greatly within bacterial colonies, between $0.6-0.8 \times 10^{-4}$ s$^{-1}$ for rapidly growing bacteria, to $0.1-1$ {hr}$^{-1}$, or even less for slowly growing ones. The slowdown of swim speed with density can be driven either by quorum sensing between the microorganisms, or by steric effects. The former is probably more likely in bacterial suspensions, as typically the equilibrium density that they would grow to, $\rho_0$, which sets a scale for typical values of $\rho$, is below 1\%, at which it is unlikely that steric interactions play a major role.

Eq.~(\ref{logistic}) leads to an arrest of the phase separation,
due to the additional reproduction (and death) time scale which is set up
by the logistic term, $\alpha\rho\left(1-\frac{\rho}{\rho_0}\right)$.
The mechanism leading to pattern formation and selection is as follows.
First, MIPS leads to the formation of clusters, or equivalently to an
inverse diffusive flux which drives bacteria, or active particles, from dilute 
regions into particle-rich ones. This flux is counteracted by the logistic
term, which tends to decrease the local density if it is above the target value
$\rho_0$ (indeed $\rho\equiv \rho_0$ is the only stable solution for 
bacteria with a swim speed which is constant, or which decreases only mildly
with density~\cite{Cates10}). The competition between these two terms
leads to a steady state where the bacteria accumulate only to a certain point,
and where multiple clusters coexist in steady state. One important finite
length scale, which is to a good approximation related to the intercluster
mean separation in steady state, is given by 
$\sqrt{{\mathcal D}/\alpha}$, where ${\mathcal D}$ is an effective
diffusion constant (depending on the density imbalance between the particle-rich
and the dilute phase). Only if $\alpha\to 0$ (no reproduction), 
coarsening can proceed indefinitely, corresponding to macroscopic phase
separation discussed in Sections 2 and 3. 

\begin{figure}[!ht]
\includegraphics[width=1.0\textwidth]{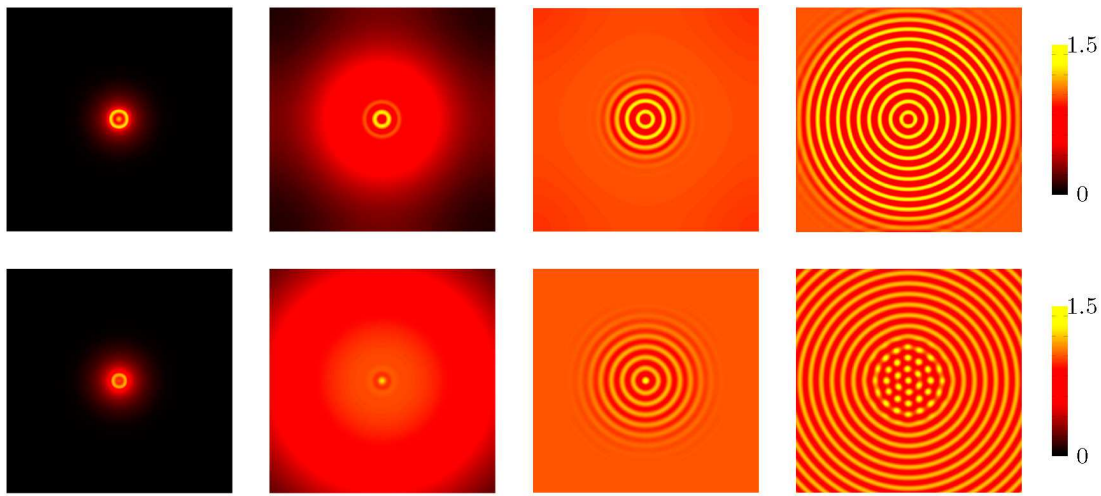}
\caption{Patterns formed in a colony of self-reproducing bacteria
with density dependent swim speed; reproduction and death are modelled
by a logistic growth law. The initial condition was an inoculum
(small bacterial droplet at the centre of the simulation box).
Each row provides a series of snapshots observed for increasing time;
the color refers to bacterial density (see colorbar on the right).
More details on parameter values etc. are found in Ref.~\cite{Cates10}.
Figure taken from~\cite{Cates10}, with permission.}
\label{fig:PNAS}
\end{figure}

The resulting pattern depends strongly on the initial
condition. For instance, starting from a uniform phase with noise leads to the
formation of droplets, whereas starting from a bacterial droplet, or
``inoculum'' (i.e. with the density concentrated in a small region initially)
leads to the formation of concentric rings or dot patterns, not unlike those
which are usually associated with chemotactic bacteria~\cite{Murray}
(see Fig.~\ref{fig:PNAS}).

Interestingly, there are other experiments which suggest that, even with synthetic self-propelled particles, motility-induced phase separation results in separate clusters rather than in macroscopic phase separation~\cite{Palacci10a,Palacci13,Palacci14}. The mechanism leading to arrest of coarsening under those conditions is still unclear. Clearly, it cannot be the one discussed previously, because while the overall density of synthetic self-propelled particles is conserved in the experiments of Ref.~\cite{Palacci10a,Palacci13,Palacci14}, it is not in Eq.~(\ref{logistic}), or in the bacterial pattern experiments it relates to.

\subsection{The role of solvent-mediated hydrodynamic interactions}
  
Up until now we have described models of ``dry'' active matter~\cite{Marchetti13}, i.e. where the solvent where active colloids typically move is neglected. In reality, microscopic swimmers (which move at low Reynolds numbers) stir the fluid they are in, hence they interact each other also through the solvent hydrodynamic flow field. This important fact, not considered in the previous Sections,  can have relevant effects on the phase behavior. We will now briefly discuss this point in this Section. 
Models for active matter under fixed  flow profile have been considered in~\cite{Saracco11,Saracco12,Rafai14}.

There are different swimming mechanisms used by microorganisms, artificial active colloids and active droplets. Here we consider the most studied case of the so-called "squirmers" that propel themselves by a prescribed axisymetric surface velocity field~\cite{Lighthill52,Blake71}.
They can be used as a simplified generic model for diffusiophoretic particles~\cite{Howse07,Erbe08}, or for biological microswimmers where the action of cilia or flagella can be described at mesoscopic scales by the squirmer surface velocity field~\cite{Lighthill52,Blake71}.      
A  review on the role of hydrodynamics in the  self-propulsion mechanisms is given in~\cite{Lauga09}.
Here, we will focus on the effects of  the hydrodynamic interactions on the collective behavior of  squirmers and, in particular, on the possibility of having a (motility-induced) phase separation as in the cases discussed in the previous sections. 

\begin{figure}[!ht]
\includegraphics[width=0.9\textwidth]{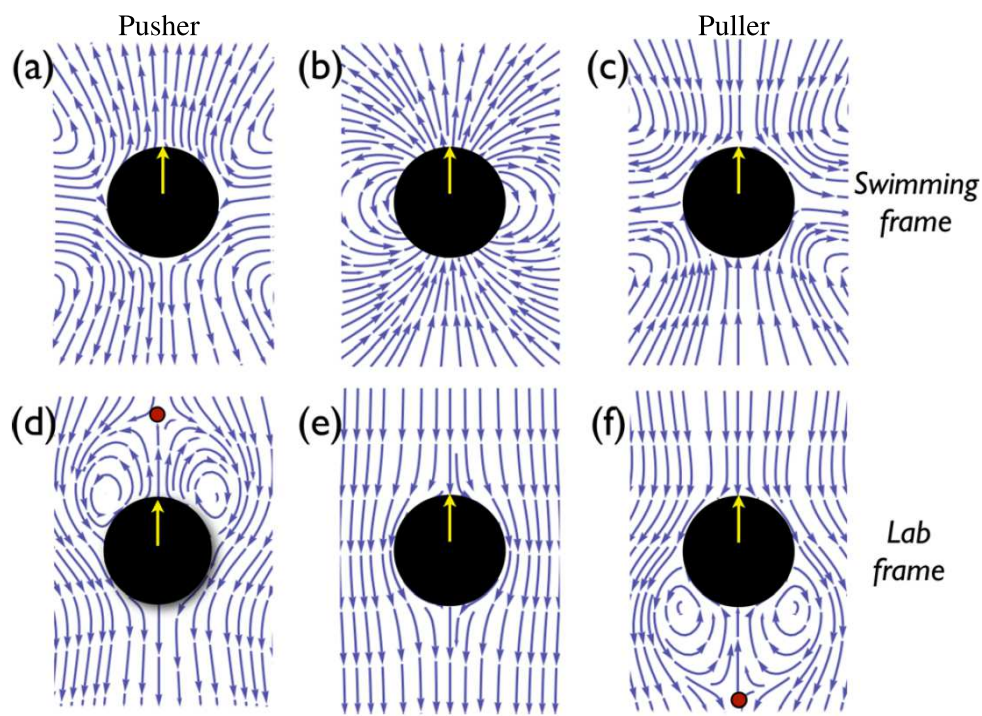}
\caption{Flow streamlines generated from an isolated squirmer in the swimming frame (top row)
and in the lab framce (bottom row). Left and right panels ((a)-(d) and (c)-(f)) show a pusher
and a puller, respectively, and the corresponding dipolar velocity field. The middle panel
((b)-(e)) shows the quadrupolar velocity field of a neutral swimmer. The figure is taken form Ref.~\cite{Evans11}.} 
\label{fig:velocityfield_squirmers}
\end{figure}

We first show in Fig.~\ref{fig:velocityfield_squirmers} the velocity field of different kinds of single squirmers.
In order to describe different real microswimmers,  the characteristic of the model can be varied to describe either pullers (equivalently contractile swimmers), with a vortex behind the body (Fig.~\ref{fig:velocityfield_squirmers}(c),(f)), or pushers (equivalently extensile swimmers) with the vortex in front of the body (Fig.~\ref{fig:velocityfield_squirmers}(a),(d)).
Fig.~\ref{fig:velocityfield_squirmers} also shows the velocity field for the case of neutral swimmers. 
The velocity field associated with a moving puller or pusher is dipolar (the velocity of the solvent decays with distance from the swimmer, $r$, as $r^{-2}$), whereas that associated with a neutral swimmer is quadrupolar (the velocity decays as $r^{-3}$).

The simplest way to analyze the hydrodynamic  interaction between squirmers
is to consider the flow field around a pair of squirmers kept at fixed position~\cite{Gotze10}.
In the case of pullers, 
%as it appears clear from Fig.~\ref{fig:velocityfield_squirmers}, 
 the fluid is pressed into the region between the two squirmers 
increasing the pressure there, and leading to repulsion. 
On the other hand, in the case of pushers,  the squirmers are expected to attract each other.
An analytic expression for the force between two squirmers is available decaying  with the inverse of the square of the distance between
the squirmers, at large distances~\cite{Ishikawa06,Blake71}. 
However, other effects can combine with the main hydrodynamical interaction, rendering the dynamics of two squirmers more complex. 
For example, thermal fluctuations may play a relevant role~\cite{Gotze10}.
Pushers initially oriented in parallel, allowed to swim freely, are observed to move towards each other but, due to diffusion, they do not always collide~\cite{Gotze10}. Sometimes, after collision, rotational diffusion reorients the squirmers and the trajectories diverge, with an  increasing average  spreading with decreasing P\'eclet number. 
 
There are few studies of  the collective behavior of squirmers. 
Simulations based on Stokesian dynamics~\cite{Ishikawa08,Ishikawa06} reported in ~\cite{Evans11} showed that pullers are systematically more ordered than pusher spherical suspensions. In this work, order was measured in terms
 of the global polarization value coming from the average of the orientation 
vectors of all particles.
In some way, this result contradicts the  expectation based on the attractive (repulsive) character of  the interaction between pushers (pullers).  This behavior was attributed to a prevalence of head-to-head collisions favouring a faster decorrelation of  pusher swimmers. 
%with 
The result was also shown to be independent on the initial, aligned or isotropic configuration of the ensemble of particles.
It also  contrasts a prediction from continuum theories saying that an isotropic  state is unstable only for nonspherical pushers~\cite{Saintillan08}. The number of particles considered in the simulations of~\cite{Evans11} was, however, too small in order to properly discuss the possibility of a phase transition. 

A mixed simulation method, with a Lattice Boltzmann scheme for the fluid coupled to particle dynamics for the squirmers 
was considered  in \cite{Alarcon13}. Here the number of particles was much larger, few thousands, and the system was 3D.
It was  shown that pullers have a tendency to form transient aggregates. This was related to the presence of polar order, which occurs for pullers but not in the case of pushers. However, a macroscopic motility-induced phase separation was not observed.  The origin of the polar order was attributed to a mechanism similar to that discussed in~\cite{Evans11}.
The aggregation phenomenon was shown to be almost absent in the case of pushers.

Using simulations methods similar to those of~\cite{Evans11}, it was shown in~\cite{Fielding14} that hydrodynamics should suppress phase separation for isometric squirmers in 2D; this was explained by mapping the model to a system of active particle in a range of parameters where motility-induced phase separation does not occur. In this work, most of the simulations were done with neutral swimmers (although simulations suggested that the results were similar for pushers or pullers).

The same squirmer system was considered in another recent work by Z{\"o}ttl and Stark~\cite{Zottl14}, that presents the results of simulations based on the multiparticle collisional dynamics method coupled to particle dynamics for the squirmers. The  swimmers move between two walls, which allow both 2D translations and full 3D rotations.
Unlike Ref.~\cite{Fielding14}, this study found a pronounced difference in behaviour in the two cases. The main result of this work is that at sufficient high area fraction $\phi_0 \ge 0.5$ the system can separate in a gaslike and a crystalline phase, depending on the nature of the swimmers. Pusher are observed to form a single cluster at $\phi_0 \approx 0.65$ while pullers are observed to form several exagonal structures and a single cluster only at very high density.  On the other hand, aggregation in clusters is seen to be  greatly favoured for neutral swimmers and diminishes when the pusher character increases. These features are explained in terms of a  strong enhancement of the rotational diffusion coefficient, compared with that of a single swimmer, due to hydrodynamic swimmmer-swimmer and swimmer-wall interactions. 

Finally, we mention the work in Ref.~\cite{Furukawa14}, which used a fluid-particle dynamics approach, and studied the effect of hydrodynamics on the collective dynamics of dumbbell swimmers with prescribed dipolar force pairs in 3D. This work showed that within semidilute suspensions of swimmers (in the presence of thermal fluctuations) hydrodynamic interactions appear to enhance the dynamic clustering (which can be seen as a precursor of fully-developed MIPS) at relatively small volume fractions. 
Within this model, this result arises due to hydrodynamic trapping of one swimmer by another, induced by the active forces.
The final prediction of this model regarding MIPS is still pending. Preliminary simulations within this framework suggest that motility-induced phase separation should occur for sufficiently large propulsion forces (at a volume fractions of $0.2-0.25$); however the structure formed was not compact and exhibited large
fluctuations, possibly because of the relatively small system size considered. A final assessment of the fate of MIPS in this model therefore requires even larger scale simulations.

As is clear from the above, gaining an ultimate understanding of the role of hydrodynamics interactions in motility-induced phase separation is a highly non-trivial, and still largely open question. This is due both to the fact that near field effects, which are difficult to capture accurately, seem to matter, and to the need to run very large scale simulations (as in the case of ABP~\cite{Stenhammar13,Stenhammar14}) to come to a firm conclusion on phase separation in the bulk. In this sense, large scale simulations in the future may hold the key to finding the answer to this difficult question. 

\section{Conclusions}

In conclusion, we have reviewed here the physics and dynamics of motility-induced phase separation, a nonequilibrium phenomenon which occurs in a suspension of self-propelled active particles. Motility-induced phase separation was first predicted theoretically in~\cite{Tailleur08} (a nice review on the topic, complementary to our work here, is the recent Ref.~\cite{Cates14}). The physics underlying this phenomenon can be understood as follows. Imagine that due to a fluctuation the density of the suspension increases locally: because self-propelled particles (unlike Brownian colloidal particles) accumulate where they are slow (see Section 2 and Refs. therein), then this fluctuation can trigger a positive feedback loop whereby particles accumulate further, slow down even more etc, eventually resulting in macroscopic phase separation, at least for suitable values of the density and propulsion speed (or P\'eclet number). Microscopically, the formation of dense regions can be traced down to the self-trapping of active particles, which move much faster than their diffusion, hence bump into each other to form clusters from which they cannot easily escape.

The physics and dynamics of motility-induced phase separation shares several aspects with that of gas-liquid systems, or binary mixtures. Importantly, the baseline model predicts that the particle-rich domains should coarsen, with a growth law which is compatible with that of a diffusive Cahn-Hilliard equation, both in 2D and 3D. This conclusion is gained, for instance, from an analysis of very large scale Brownian dynamics simulations of spherical active Brownian particles, and is also confirmed in 2D lattice simulations of run-and-tumble particles. We have also seen that the phenomenology of motility-induced phase separation has some important differences when considering elongated particles, such as dumbbells. Here, particle-based simulations show that the aggregates which form due to the self-trapping mechanism rotate coherently, and we have seen how the dependence of angular velocity on cluster size can be understood with simple scaling arguments. 

Importantly, coarsening can be arrested; an intriguing example is provided by a population of reproducing run-and-tumble particles with a density-dependent swim speed, which would lead, in the absence of reproduction, to macroscopic phase separation. The presence of an additional time scale, linked to reproduction and death of bacteria, which can be, for instance, captured by a simple logistic law, allows phase separation to be arrested. The resulting patterns are very similar to those found in bacterial colonies growing in semisolid media, such as agarose gel. Such patterns were normally attributed to chemotaxis; the results in Section 4 suggest that arrested motility-induced phase separation provides another viable mechanism for these clusters.

There are several open questions which are needed to fully understand motility-induced phase separation, and we have mentioned some of these in our work. First, it is unclear to what extent the diffusive scaling exhibited by spherical self-propelled particles, and the associated continuum theory, applies to suspensions of anisometric active particles, such as dumbbells or rods. Second, a few of the experiments studying the phase behaviour of self-propelled particles~\cite{Palacci13,Bocquet} seem to be consistent with arrested phase separation. The possible mechanism leading to arrest of coarsening in this context is unclear at the moment. Third, the role of solvent-mediated hydrodynamic interactions are currently not fully understood; yet the simulations performed to date suggest that they are likely to affect the physics of motility-induced phase separation significantly. We hope that our review will stimulate further work on this interesting nonequilibrium phase transition in active matter.

%\vspace{0.25cm}

%\noindent \underline{Acknowledgments}:
%  G.G. acknowledges the support of  MIUR (project PRIN 2012NNRKAF).

\newpage

\bibliographystyle{elsarticle-num} 
\bibliography{mips_review}

\end{document}